\documentclass[preprint,nofootinbib]{revtex4}%
\usepackage{amssymb}
\usepackage{amsfonts}
\usepackage{amsmath}
\usepackage{graphicx}%
\providecommand{\U}[1]{\protect\rule{.1in}{.1in}}
\begin{document}
\title{Gravitational solitons, hairy black holes \\and phase transitions in BHT massive gravity}
\author{Alfredo P\'{e}rez, David Tempo, Ricardo Troncoso}
\affiliation{Centro de Estudios Cient\'{\i}ficos (CECs), Casilla 1469, Valdivia, Chile.}
\preprint{CECS-PHY-10/14}

\begin{abstract}
Hairy black holes and gravitational solitons in three dimensions for the new
massive gravity theory proposed by Bergshoeff, Hohm and Townsend (BHT) are
considered at the special case when there is a unique maximally symmetric
solution.\textbf{ }Following the Brown-York approach with suitable
counterterms, it is shown that the soliton possesses a fixed negative mass
which coincides with the one of AdS spacetime regardless the value of the
integration constant that describes it. Hence, the soliton can be naturally
regarded as a degenerate ground state labeled by a single modulus parameter.
The Euclidean action, endowed with suitable counterterms, is shown to be
finite and independent of modulus and hair parameters for both classes of
solutions, and in the case of hairy black holes the free energy in the
semiclassical approximation is reproduced. Modular invariance allows to show
that the gravitational hair turns out to be determined by the modulus
parameter. According to Cardy's formula, it is shown that the semiclassical
entropy agrees with the microscopic counting of states provided the modulus
parameter of the ground state is spontaneously fixed, which suggests that the
hairy black hole is in a broken phase of the theory. Indeed,  it is found that
there is a critical temperature $T_{c}=(2\pi l)^{-1}$ characterizing a first
order phase transition between the static hairy black hole and the soliton
which, due to the existence of gravitational hair, can take place in the
semiclassical regime.

\end{abstract}
\maketitle

\section{Introduction}

Three-dimensional gravity has recently received a great deal of attention,
specially in the case of extensions of General Relativity that admit
propagating massive gravitons. Apart from the well-known theory of
topologically massive gravity \cite{TMG1,TMG2}, a different theory currently
known as \textquotedblleft new massive gravity\textquotedblright\ has been
proposed by Bergshoeff, Hohm and Townsend (BHT) \cite{BHT,BHT2}. The theory is
described by the following action%
\begin{equation}
I_{_{\text{\textrm{BHT}}}}=\frac{1}{16\pi G}\int_{M}d^{3}x\sqrt{-g}\left[
R-2\lambda-\frac{1}{m^{2}}\left(  R_{\mu\nu}R^{\mu\nu}-\frac{3}{8}%
R^{2}\right)  \right]  \ ,\label{IBHT}%
\end{equation}
which is manifestly invariant under parity and gives fourth order field
equations for the metric. Remarkably, at the linearized level the field
equations reduce to the ones of Fierz and Pauli for a massive graviton,
generically describing two independent propagating degrees of freedom. This
has also been confirmed at the full nonlinear level from different canonical
approaches \cite{Hamiltonian1,Hamiltonian1.1,Hamiltonian2}. A wide variety of
exact solutions can be found in Refs.
\cite{OTT,BHT2,Dynamics-BHT,WarpedBH-Solution,1NullKilling-Solution,TypeD-Solutions,TypeN-Solutions,LifshitzBH,ABGH,Kundt-Spacetimes,Bounce-Solutions,DomainWall-Solutions,Homogeneous-Solutions,ChargedBH-Solutions}%
, and further aspects of the theory have been developed in
\cite{QNormalModes-BHT,C-Functions-GulluSismanTekin,NewAnomalies,C-theorem-Sinha,Sinha-AdS/CFT}%
. Hereafter we will focus in the special case
\begin{equation}
m^{2}=\lambda\ ,\label{special point}%
\end{equation}
since the theory possesses additional interesting features. In this case the
field equations admit a unique maximally symmetric solution (with
$R_{\ \alpha\beta}^{\mu\nu}=\Lambda\delta_{\alpha\beta}^{\mu\nu}$) whose
curvature radius is determined by $\Lambda=2m^{2}$, and at the linearized
level the graviton becomes \textquotedblleft partially
massless\textquotedblright\ \cite{BHT2,DeserALAS,PM1,PM2,PM3,PM4}. The special
case (\ref{special point}) enjoys further remarkable properties, as it is the
existence of interesting solutions including hairy black holes and
gravitational solitons \cite{OTT}. Hereafter we will focus on these latter
solutions in the case of negative cosmological constant $\Lambda:=-\frac
{1}{l^{2}}$. We show that the soliton possesses a negative fixed mass which
agrees with that of AdS, so that its integration constant turns out to be a
modulus parameter. The regularized Euclidean action becomes independent of
modulus and hair parameters for both classes of solutions, and it reduces to
the free energy in the case of hairy black holes. An interesting link between
both solutions can be seen through modular invariance, which allows to find a
precise relationship between the gravitational hair of the black hole and the
modulus parameter of the soliton. Noteworthy, it is found that this relation
exactly maps the mass bound for the hairy black hole required by cosmic
censorship with the condition that guarantees the regularity of the soliton.
Furthermore, the semiclassical entropy agrees with the microscopic counting of
states according to Cardy formula provided the modulus parameter of the ground
state is spontaneously fixed, which suggests that the hairy black hole is in a
broken phase of the theory. Indeed, it is found that there is a critical
temperature that characterizes a first order phase transition between the
static hairy black hole and the soliton. Remarkably, the presence of
gravitational hair induces an additional effective length scale which allows a
suitable treatment of this phase transition in the semiclassical regime.

Our paper is organized as follows. In the next section the gravitational
soliton is briefly reviewed, while its mass is obtained through the Brown-York
approach with suitable counterterms in Sec.
\ref{Soliton mass + Euclidean action}. The regularized Euclidean action is
also analyzed in the case of the rotating hairy black hole in Sec.
\ref{Rotating black hole + Euclidean action}, where it is shown to be
independent of the hair parameter, and reduces to the free energy in the
semiclassical approximation. Modular invariance is discussed in Sec.
\ref{Modular invariance section}, which allows to show that the gravitational
hair of the back hole turns out to be determined by the modulus parameter of
the soliton. In Sec. \ref{Cardy} it is shown that the semiclassical entropy
agrees with the microscopic counting of states provided the ground state is
non degenerate. As discussed in Sec.\ref{HBHPT} a first order phase transition
between the static hairy black hole and the soliton is shown to occur at a
critical temperature $T_{c}=(2\pi l)^{-1}$, while in Sec.
\ref{Thermal fluctuations} we explain how the existence of gravitational hair
may allow the transition to take place in the semiclassical regime. Finally,
Sec. \ref{Discussion} is devoted to the summary and discussion.

\section{Soliton as a degenerate ground state}

\label{Soliton}

It has been shown that the field equations that correspond to the action
(\ref{IBHT}) at the special point (\ref{special point}), in the case of
negative cosmological constant admit a solution whose metric is given by
\cite{OTT}%
\begin{equation}
ds^{2}=l^{2}\left[  -\left(  \alpha+\cosh\rho\right)  ^{2}dt^{2}+d\rho
^{2}+\sinh^{2}\rho d\varphi^{2}\right]  \ . \label{Soliton nice coordinates}%
\end{equation}
If the integration constant $\alpha$ fulfills
\begin{equation}
\alpha>-1\ , \label{Soliton existence condition}%
\end{equation}
this spacetime is smooth and regular everywhere and describes a gravitational
soliton. Note that in the case of $\alpha=0$, the soliton
(\ref{Soliton nice coordinates}) reduces to AdS spacetime in global
coordinates. The soliton can then be regarded as a smooth deformation of AdS
spacetime, sharing the same causal structure. Furthermore, the metric
(\ref{Soliton nice coordinates}) is asymptotically AdS in a relaxed sense as
compared with the one of Brown and Henneaux \cite{Brown-Henneaux}, but
nevertheless the asymptotic symmetries remains the same. Indeed, the soliton
can be written in Schwarzschild-like coordinates making%
\begin{equation}
\sinh\rho\rightarrow\frac{r}{l}\ ;\ t\rightarrow\frac{t}{l}\ ,
\end{equation}
so that the metric reads%

\begin{equation}
ds^{2}=-\left(  \alpha+\sqrt{\frac{r^{2}}{l^{2}}+1}\right)  ^{2}dt^{2}%
+\frac{dr^{2}}{\frac{r^{2}}{l^{2}}+1}+r^{2}d\varphi^{2}\ ,
\label{Soliton-Scharzschild-like}%
\end{equation}
and the deviation with respect to the AdS metric at infinity is of the form
$\Delta g_{tt}=-\frac{2\alpha}{l}r+\mathcal{O}(1)$. Note that the deviation
grows instead of decaying as one approaches to the asymptotic region since it
is linearly divergent, in sharp contrast with the standard behavior, given by
$\Delta g_{tt}=\mathcal{O}(1)$ \cite{Brown-Henneaux}. In spite of this
divergent behavior, finite charges as surface integrals can be constructed
through standard perturbative methods \cite{Abbott-Deser,Deser-Tekin}.
However, in this case, quadratic deviations with respect to the background
metric turn out to be relevant since they give additional nontrivial
contributions to the surface integrals. The purpose of the next subsection is
to compute the mass of the soliton within a fully nonlinear approach.

\subsection{Soliton mass and Euclidean action from the Brown-York approach
with suitable counterterms}

\label{Soliton mass + Euclidean action}

Here it is shown that the soliton (\ref{Soliton nice coordinates}) possesses a
negative fixed mass which coincides with the one of AdS spacetime, so that the
integration constant $\alpha$ corresponds to a modulus parameter. The approach
we follow is the quasilocal\ one of Brown and York \cite{Brown-York}, where
the action is regularized with suitable counterterms along the lines of
\cite{Henningson-Skenderis, Balasubramanian-Kraus}. For the BHT\ massive
gravity theory, this task was performed by Hohm and Tonni \cite{Hohm-Tonni}
for a generic value of the parameter $m^{2}$, and in the special case
(\ref{special point}) by Giribet and Leston \cite{Giribet-Leston}. Thus, in
our case, the suitable regularized action turns out be given by
\begin{equation}
I_{\text{\textrm{reg}}}=I_{_{\text{\textrm{BHT}}}}+I_{_{\text{\textrm{GH}}}%
}+I_{\text{\textrm{ct}}}\ .\label{reg1-1}%
\end{equation}
It is useful expressing the bulk action in second-order form by means of the
auxiliary field $f^{\mu\nu}$, so that it reads%
\begin{equation}
I_{_{\text{\textrm{BHT}}}}=\frac{1}{16\pi G}\int_{M}d^{3}x\sqrt{-g}\left[
R-2\lambda-f^{\mu\nu}G_{\mu\nu}+\frac{1}{4}m^{2}\left(  f_{\mu\nu}f^{\mu\nu
}-f^{2}\right)  \right]  \ ,\label{reg2}%
\end{equation}
and the boundary terms can be written as%
\begin{align}
I_{_{\text{\textrm{GH}}}} &  =\frac{1}{16\pi G}\int_{\partial M}d^{2}%
x\sqrt{-\gamma}\left[  -2K-\hat{f^{ij}}K_{ij}+\hat{f}K\right]  \ ,\label{reg3}%
\\
I_{\text{\textrm{ct}}} &  =\frac{1}{16\pi Gl}\int_{\partial M}d^{2}%
x\sqrt{-\gamma}\hat{f}\ ,\label{reg4}%
\end{align}
where $\hat{f^{ij}}$ is defined in terms of $f^{ij}$ and the shift $N^{j}$ of
a radial ADM decomposition of the bulk metric according to%
\begin{equation}
\hat{f^{ij}}=f^{ij}+2f^{r(i}N^{j)}+f^{rr}N^{i}N^{j}\ .
\end{equation}
It is also useful to express the boundary metric of $\partial M$ in the
standard ADM foliation with spacelike surfaces $\Sigma$, i.e.,%
\begin{equation}
\gamma_{ij}dx^{i}dx^{j}=-N_{\Sigma}^{2}dt^{2}+\sigma(d\varphi+N_{\Sigma
}^{\varphi}dt)(d\varphi+N_{\Sigma}^{\varphi}dt)\ ,
\end{equation}
so that the Brown-York stress tensor \cite{Brown-York} is obtained by varying
the regularized action with respect to the boundary metric $\gamma_{ij}$%
\begin{equation}
T^{ij}=\lim_{r\rightarrow\infty}\frac{2}{\sqrt{-\gamma}}\frac{\delta
I_{\text{\textrm{reg}}}}{\delta\gamma_{ij}}\ .\label{Tij}%
\end{equation}
Therefore, the corresponding conserved charge associated to a Killing vector
$\xi$ is given by%
\begin{equation}
Q(\xi)=\int_{\Sigma}d\varphi\sqrt{\sigma}u^{i}\xi^{j}T_{ij}\ ,
\end{equation}
where $u^{i}$ is the timelike unit normal to $\Sigma$.

In the case of the soliton metric (\ref{Soliton nice coordinates}), the
Brown-York stress-energy tensor reads%
\begin{equation}
T_{ij}^{_{\mathrm{sol}}}=\left(
\begin{array}
[c]{cc}%
-\frac{1}{8\pi lG} & 0\\
0 & -\frac{l}{8\pi G}%
\end{array}
\right)  \ ,
\end{equation}
and hence, its mass is finite, negative, and given by%
\begin{equation}
M_{\mathrm{sol}}=Q(\partial_{t})=-\frac{1}{4G}\ . \label{Msol}%
\end{equation}
Note that since the soliton mass does not depend on the parameter $\alpha$, it
exactly coincides with the one of AdS spacetime. Therefore, as this
integration constant plays no role in the conserved charges, the soliton can
be naturally regarded as a degenerate ground state labeled by a single modulus parameter.

\bigskip

The Euclidean continuation of the soliton can be easily obtained through
$t\rightarrow i\tau_{E}$, where $\tau_{E}$ is the \textquotedblleft Euclidean
time\textquotedblright\ with an arbitrary period $\beta$, and since it
possesses finite Euclidean action it describes an instanton. This can be
explicitly seen as follows. Evaluating the Euclidean continuation of
(\ref{Soliton-Scharzschild-like}) one obtains that the relevant terms are
given by%
\begin{equation}
I_{_{\text{\textrm{BHT}}}}=-\frac{\beta}{2G}\left(  \frac{r^{2}}{l^{2}%
}\right)  \ ,\ I_{_{\text{\textrm{GH}}}}=\frac{\beta}{G}\left(  \frac{r^{2}%
}{l^{2}}+\frac{1}{2}\right)  \ ,\ I_{\text{\textrm{ct}}}=-\frac{\beta}%
{2G}\left(  \frac{r^{2}}{l^{2}}+\frac{1}{2}\right)  \ ,
\end{equation}
so that the divergences exactly cancel out and the total Euclidean action
(\ref{reg1-1}) for the soliton becomes finite and given by%
\begin{equation}
I_{\text{\textrm{sol}}}=-\beta M_{\text{\textrm{sol}}}=\frac{1}{4G}\beta\ .
\label{Isol}%
\end{equation}

As expected, the Euclidean action does not depend on the modulus parameter
$\alpha$, and therefore it coincides with that of AdS spacetime. As a cross
check, the soliton mass can alternatively be computed according to
\begin{equation}
M_{\mathrm{sol}}=-\partial_{\beta}I_{\text{\textrm{reg}}}=-\frac{1}%
{4G}\ ,\label{Msol2}%
\end{equation}
which precisely agrees with the result found above in (\ref{Msol}) by means of
the Brown-York stress-energy tensor.

\section{Rotating hairy black hole: regularized Euclidean action and
thermodynamics}

\label{Rotating black hole + Euclidean action}

The regularized Euclidean action (\ref{reg1-1}) of the rotating hairy black
hole turns out to be finite and it reduces to the free energy in the
semiclassical approximation. As naturally expected, it depends only on the
Hawking temperature and the angular velocity of the horizon, $\beta$ and
$\Omega_{+}$, respectively, and it is then independent of the gravitational
hair parameter. This can be seen as follows: The field equations at the
special point (\ref{special point}) admit a solution whose metric is given by
\cite{OTT}%

\begin{equation}
ds^{2}=-G\left(  r\right)  dt^{2}+\frac{dr^{2}}{F\left(  r\right)  }+2N^{\phi
}\left(  r\right)  dtd\phi+\left(  r^{2}+r_{0}^{2}\right)  d\phi^{2}\ ,
\label{Rot-BH}%
\end{equation}
where%
\begin{align}
F\left(  r\right)   &  :=\frac{r^{2}}{l^{2}}-\frac{b}{l}\left(  1+\Xi
^{\frac{-1}{2}}\right)  r+\frac{b^{2}}{4}\left(  1+\Xi^{\frac{-1}{2}}\right)
^{2}-4G\mathcal{M}\Xi^{\frac{1}{2}}\ ,\\
G\left(  r\right)   &  :=\frac{r^{2}}{l^{2}}-2\frac{b}{l}\Xi^{\frac{-1}{2}%
}r+\frac{b^{2}}{4}\left(  1+3\Xi^{-1}\right)  -2G\mathcal{M}\left(
1+\Xi^{\frac{1}{2}}\right)  \ ,\\
N^{\phi}\left(  r\right)   &  :=-a\left(  \frac{b}{l}\Xi^{\frac{-1}{2}%
}r+2G\mathcal{M}+\frac{b^{2}}{2}\Xi^{-1}\right)  \ ,
\end{align}
and the constants $r_{o}^{2}$, and $\Xi$ are defined as%

\begin{align*}
r_{0}^{2}  &  :=\frac{b^{2}l^{2}}{4}\left(  1-\Xi^{-1}\right)  +2G\mathcal{M}%
l^{2}\left(  1-\Xi^{\frac{1}{2}}\right)  \ ,\\
\Xi &  :=1-\frac{a^{2}}{l^{2}}\ ,
\end{align*}
The solution depends on three integration constants, where $\mathcal{M}$
corresponds to the mass, the angular momentum is given by $\mathcal{J}%
=\mathcal{M}a$ (with $a^{2}\leq l^{2}$), and $b$ is the gravitational hair
parameter\footnote{For simplicity, here the gravitational hair parameter $b$
has been redefined making $b\rightarrow-2bl^{-1}$ in \cite{OTT}.}. Note that
the BTZ black hole \cite{BTZ,BHTZ}\ is recovered for $b=0$. The singularity at
$r=r_{s}:=-\frac{1}{2}bl\left(  1-\Xi^{\frac{-1}{2}}\right)  $ is cloaked by
the event horizon located at $r=r_{+}$, with%
\begin{equation}
r_{+}=\frac{bl}{2}\left(  1+\Xi^{-\frac{1}{2}}\right)  +2l\sqrt{G\mathcal{M}%
\Xi^{\frac{1}{2}}}\ .
\end{equation}
Cosmic censorship then requires that $r_{+}\geq r_{s}$, which in the case of
negative $b$, implies the following bound for the mass%
\begin{equation}
\mathcal{M}\geq\frac{b^{2}\Xi^{-\frac{1}{2}}}{4G}\ , \label{Mass bound}%
\end{equation}
while for positive $b$ the bound is such that mass turns out to be nonnegative.

The angular velocity of the horizon is given by%
\[
\Omega_{+}=\frac{1}{a}\left(  \Xi^{\frac{1}{2}}-1\right)  \ ,
\]
and the Hawking temperature expressed in terms of the Euclidean time period,
$T=\beta^{-1}$, reads%
\begin{equation}
\beta^{2}=\frac{\pi^{2}l^{2}}{2\mathcal{M}G}\left(  1+\Xi^{\frac{1}{2}%
}\right)  \Xi^{-1}\ , \label{Beta-Temperature}%
\end{equation}
where the Euclidean continuation of the rotating black hole is performed
through $t\rightarrow it_{E}$, and $a\rightarrow ia$.

Evaluating the Euclidean rotating hairy black hole on each term of the
regularized action (\ref{reg1-1}), one obtains%
\begin{align*}
I_{_{\text{\textrm{BHT}}}} &  =-\frac{\beta}{2G}\left(  \frac{r^{2}}{l^{2}%
}-\frac{b}{l}r\left(  1+\Xi^{-\frac{1}{2}}\right)  +\frac{b^{2}}{4}\left(
1+\Xi^{\frac{1}{2}}\right)  ^{2}\Xi^{-1}-4G\mathcal{M}\Xi^{\frac{1}{2}%
}\right)  \ ,\\
I_{_{\text{\textrm{GH}}}} &  =\frac{\beta}{G}\left(  \frac{r^{2}}{l^{2}}%
-\frac{b}{l}r\left(  1+\Xi^{-\frac{1}{2}}\right)  +\frac{b^{2}}{4}\left(
1+\Xi^{\frac{1}{2}}\right)  ^{2}\Xi^{-1}-2G\mathcal{M}\Xi^{\frac{1}{2}%
}\right)  \ ,\\
I_{\text{\textrm{ct}}} &  =-\frac{\beta}{2G}\left(  \frac{r^{2}}{l^{2}}%
-\frac{b}{l}r\left(  1+\Xi^{-\frac{1}{2}}\right)  +\frac{b^{2}}{4}\left(
1+\Xi^{\frac{1}{2}}\right)  ^{2}\Xi^{-1}-2G\mathcal{M}\Xi^{\frac{1}{2}%
}\right)  \ .
\end{align*}
Hence, the divergent terms cancel out so that the total Euclidean action
(\ref{reg1-1}) of the rotating hairy black hole becomes finite and given by%
\begin{equation}
I_{\text{\textrm{hbh}}}=\beta\mathcal{M}\Xi^{\frac{1}{2}}\ .\label{IBH1}%
\end{equation}
As explained in \cite{OTT, GOTT}, since the gravitational hair parameter $b$
is not related to a global charge, no chemical potential can be associated
with. This can be independently confirmed by virtue of Eq. (\ref{IBH1}) since
once expressed in terms of the non extensive variables $\beta$ and $\Omega
_{+}$, it reads%
\begin{equation}
I_{\text{\textrm{hbh}}}=\frac{\pi^{2}l^{2}}{G}\frac{1}{\beta}\left(  \frac
{1}{1+\Omega_{+}^{2}l^{2}}\right)  \ ,\label{IbetaOmega}%
\end{equation}
which is manifestly independent of $b$.

The Euclidean action reduces to the free energy in the semiclassical
approximation, i.e., $I_{\text{\textrm{hbh}}}=-\beta F$, with $F=\mathcal{M}%
-TS-\Omega_{+}\mathcal{J}\ $, so that the first law is recovered requiring it
to have an extremum. Indeed, the mass and the angular momentum are given by
\begin{align}
\mathcal{M} &  =\left(  -\partial_{\beta}+\beta^{-1}\Omega_{+}\partial
_{\Omega}\right)  I_{\text{\textrm{hbh}}}\ ,\label{Mass}\\
\mathcal{J} &  =\beta^{-1}\partial_{\Omega}I_{\text{\textrm{hbh}}%
}\ ,\label{Angular momentum}%
\end{align}
which coincide with the expressions found through the evaluation of the
Brown-York stress-energy tensor in \cite{Giribet-Leston}. This is also in full
agreement with previous results \cite{GOTT}. In the static case, the mass has
also been recovered by different methods in Refs. \cite{MJ-BHT,M-BHT}.
Analogously, the black hole entropy can obtained from
\[
S=\left(  1-\beta\partial_{\beta}\right)  I_{\text{\textrm{hbh}}}\ ,
\]
which gives
\begin{equation}
S=\pi l\sqrt{\frac{2\mathcal{M}}{G}\left(  1+\Xi^{\frac{1}{2}}\right)
}\ ,\label{BH-Entropy}%
\end{equation}
and agrees with the result found in \cite{OTT, GOTT} by means of Wald's
formula \cite{Wald}. Further aspects concerning the rotating hairy black hole
thermodynamics have been studied in Refs.
\cite{CorrectedS,HawkingRadiationOTT}.

\section{Modular invariance, solitons and microscopic entropy of the rotating
hairy black hole}

\label{Modular invariance section}

It is useful to express the Euclidean action of the soliton (\ref{Isol}) and
the hairy black hole (\ref{IbetaOmega}) in terms of the modular parameter of
the torus geometry at the boundary, given by%
\begin{equation}
\tau:=\frac{i\hat{\beta}}{2\pi l}\ , \label{Modular parameter}%
\end{equation}
with $\hat{\beta}:=\beta\left(  1-i\Omega_{+}l\right)  $. The corresponding
Euclidean actions then read
\begin{equation}
I_{\mathrm{sol}}=i\pi lM_{\mathrm{sol}}\left(  \tau-\bar{\tau}\right)  \ ,
\label{IsolTau}%
\end{equation}
and%
\begin{equation}
I_{\mathrm{hbh}}=-i\pi lM_{\mathrm{sol}}\left(  \frac{1}{\tau}-\frac{1}%
{\bar{\tau}}\right)  \ , \label{IhbhTau}%
\end{equation}
respectively, where $M_{\mathrm{sol}}=-\frac{1}{4G}$ stands for the soliton
mass previously obtained in Eqs. (\ref{Msol}) and (\ref{Msol2}). Note that
(\ref{IsolTau}) and (\ref{IhbhTau}) are related by a modular transformation
given by%
\begin{equation}
\tau\rightarrow-\frac{1}{\tau}\ , \label{Modular inv}%
\end{equation}
as it is the case of Euclidean AdS and the BTZ black hole in General
Relativity \cite{Maldacena-Strominger}. It is worth pointing out that the
Euclidean action of the hairy black hole is completely determined by the
soliton mass $M_{\mathrm{sol}}$ and the modular parameter $\tau$, in full
agreement with what occurs for black holes with scalar hair and scalar
solitons in General Relativity \cite{HMTZ-2+1, CMT-Soliton}.

Remarkably, although the Euclidean actions (\ref{IsolTau}) and (\ref{IhbhTau})
do not depend neither on the modulus parameter $\alpha$ of the soliton, nor on
the gravitational hair parameter $b$ of the black hole, since both solutions
are dual under modular invariance, the gravitational hair turns out to be
determined by the quotient of $\alpha$ and $|\tau|$. This can be seen as
follows: The holographic realization of modular invariance that connects both
solutions, amounts to a coordinate transformation given by\footnote{Note that
according to (\ref{ModularTransfMetrics1}) the modular transformation
(\ref{Modular inv}) swaps the role of Euclidean time and the angular
coordinate.}%
\begin{align}
\phi &  =\frac{1}{2}\left[  \left(  \tau+\bar{\tau}\right)  \varphi+i\left(
\frac{1}{\tau}-\frac{1}{\bar{\tau}}\right)  \tau_{_{E}}\right]  \ ,\nonumber\\
t_{_{E}} &  =-\frac{l}{2}\left[  i\left(  \tau-\bar{\tau}\right)
\varphi+\left(  \frac{1}{\tau}+\frac{1}{\bar{\tau}}\right)  \tau_{_{E}%
}\right]  \ ,\label{ModularTransfMetrics1}%
\end{align}
with%
\begin{equation}
r=\frac{l}{|\tau|}\cosh\rho-\frac{bl}{4}\left(  \frac{\tau-\bar{\tau}}{|\tau
|}\right)  ^{2}\ ,\label{ModularTransfMetrics2}%
\end{equation}
which means that an Euclidean rotating hairy black hole with coordinates
$(t_{_{E}},r,\phi)$ characterized by a modular and hair parameters $\tau$ and
$b$, respectively, is diffeomorphic to the Euclidean continuation of the
soliton in Eq. (\ref{Soliton nice coordinates}), with coordinates $(\tau
_{_{E}},\rho,\varphi)$ and modulus parameter $\alpha=b|\tau|$. In turn, this
naturally suggests that a soliton with modulus parameter $\alpha$ corresponds
to the ground state of a hairy black hole whose gravitational hair parameter
is given by%
\begin{equation}
b=\frac{\alpha}{|\tau|}\ .\label{Modulus-hair-map}%
\end{equation}

It is also worth highlighting that the relationship expressed by Eq.
(\ref{Modulus-hair-map}) exactly maps the mass bound required by cosmic
censorship for the hairy black hole (\ref{Mass bound}) with the condition
(\ref{Soliton existence condition}) that guarantees the smoothness of the soliton.

\subsection{Cardy formula and spontaneous fixing of ground state modulus
parameter}

\label{Cardy}

Formula (\ref{Modulus-hair-map}) further suggests that a black hole with hair
parameter $b$ is in a broken phase of the theory in which the modulus
parameter of the ground state is spontaneously fixed. In fact, in this case,
it can be seen that semiclassical entropy agrees with the microscopic counting
of states according to Cardy formula \cite{Cardy}, as it occurs for General
Relativity \cite{Strominger}. Indeed, as emphasized in \cite{CMT-Soliton}, it
is useful to express Cardy's formula in terms of the shifted Virasoro
operators, $\tilde{L}_{0}^{\pm}:=L_{0}^{\pm}-\frac{c^{\pm}}{24}$, so that it
reads%
\begin{equation}
S=4\pi\sqrt{-\tilde{\Delta}_{0}^{+}\tilde{\Delta}^{+}}+4\pi\sqrt
{-\tilde{\Delta}_{0}^{-}\tilde{\Delta}^{-}}\ ,\label{Cardy super reloaded}%
\end{equation}
where ($\tilde{\Delta}_{0}^{\pm}$) $\tilde{\Delta}^{\pm}$ correspond to the
(lowest) eigenvalues of $\tilde{L}_{0}^{\pm}$. Thus, noteworthy, the
asymptotic growth of the number of states can be obtained only in terms of the
spectrum of $\tilde{L}_{0}^{\pm}$ without making any explicit reference to the
central charges $c^{\pm}$.

The rotating hairy black hole entropy can then be computed from
(\ref{Cardy super reloaded}) assuming that the eigenvalues of $\tilde{L}%
_{0}^{\pm}$ are given by the global charges of the black hole, i.e.,%
\begin{equation}
\tilde{\Delta}^{\pm}=\frac{1}{2}(\mathcal{M}l\pm\mathcal{J})\ ,
\label{Virasoros}%
\end{equation}
where the the ground state corresponds to the soliton
(\ref{Soliton nice coordinates}), so that the lowest eigenvalues of $\tilde
{L}_{0}^{\pm}$ are given by
\[
\tilde{\Delta}_{0}^{+}=\tilde{\Delta}_{0}^{-}=\frac{l}{2}M_{\mathrm{sol}%
}=-\frac{l}{8G}\ .
\]
Since in the case under consideration, the lowest eingenvalues $\tilde{\Delta
}_{0}^{\pm}$ can be expressed in terms of the central charges\footnote{As
shown in \cite{OTT}, at the special point (\ref{special point}) the central
charges are given by twice the value of Brown and Henneaux, i.e., $c^{+}%
=c^{-}=\frac{3l}{G}$.}, i.e., $\tilde{\Delta}_{0}^{\pm}=-\frac{c^{\pm}}{24}$,
one verifies that formula (\ref{Cardy super reloaded}) reduces to its standard
form%
\begin{equation}
S=2\pi\sqrt{\frac{c^{+}}{6}\tilde{\Delta}^{+}}+2\pi\sqrt{\frac{c^{-}}{6}%
\tilde{\Delta}^{-}}\ , \label{Cardy simple}%
\end{equation}
which, as explained in \cite{GOTT} exactly reproduces the semiclassical
entropy of the rotating hairy black hole in Eq. (\ref{BH-Entropy}).

Note that in formula (\ref{Cardy super reloaded}) it has been implicitly
assumed that the ground state is non degenerate; otherwise, the asymptotic
growth of the number of states would be given by%
\begin{equation}
\rho(\tilde{\Delta}^{\pm})=\rho(\tilde{\Delta}_{0}^{\pm})\exp\left(  4\pi
\sqrt{-\tilde{\Delta}_{0}^{\pm}\tilde{\Delta}^{\pm}}\right)
\ ,\label{Number of states}%
\end{equation}
where $\rho(\tilde{\Delta}_{0}^{\pm})$ correspond to the ground state
degeneracy (see, e.g., \cite{CarlipStates, CMT-Soliton, Logarithmic
corrections}). Therefore, since formula (\ref{Cardy super reloaded}) exactly
matches the semiclassical entropy of the hairy black hole, one obtains that
$\rho(\tilde{\Delta}_{0}^{\pm})=1$, which means that the ground state
degeneracy is removed. This can be interpreted as the fact that the modulus
parameter of the ground state is spontaneously fixed, so that the hairy black
hole is actually in a broken phase of the theory. The purpose of the next
section is to show that this is the case.

\section{Hairy black hole-Soliton phase transition}

\label{HBHPT}

According to Eqs. (\ref{IsolTau}) and (\ref{IhbhTau}), at high temperatures
the partition function is dominated by the hairy black hole, while for low
temperatures it turns out to be dominated by thermal radiation on the soliton.
In order to see whether there is a phase transition involving these objects
one has to identify different possible configurations with the same fixed
modular parameter $\tau$, i.e. one has to look for more than one configuration
at fixed temperature $\beta$ and chemical potential $\Omega_{+}$. It is simple
to verify that this only occurs in the static case, where it can be seen that
at fixed temperature there exists a phase transition between the hairy black
hole and the soliton. Indeed, according to (\ref{IbetaOmega}), in the case of
the static hairy black hole the free energy is given by%
\begin{equation}
F_{\mathrm{hbh}}=-\frac{\pi^{2}l^{2}}{G}T^{2}\ ,
\end{equation}
while for the soliton, from Eq. (\ref{Isol}), the free energy turns out to be%
\begin{equation}
F_{\text{\textrm{sol}}}=-\frac{1}{4G}\ .
\end{equation}
Therefore, at the critical temperature%
\begin{equation}
T=T_{c}:=\frac{1}{2\pi l}\ ,\label{Tc}%
\end{equation}
which corresponds to the self-dual point of the modular transformation
(\ref{Modular inv}), the soliton and the hairy black hole possess the same
free energy. This means that below the critical point ($T<T_{c}$), the soliton
has less free energy than the hairy black hole, while for $T>T_{c}$ the hairy
black hole is the configuration that dominates the partition function (See
Fig. 1).
\begin{figure}[ptb]
\begin{center}
\includegraphics[angle=-90,width=15 cm]{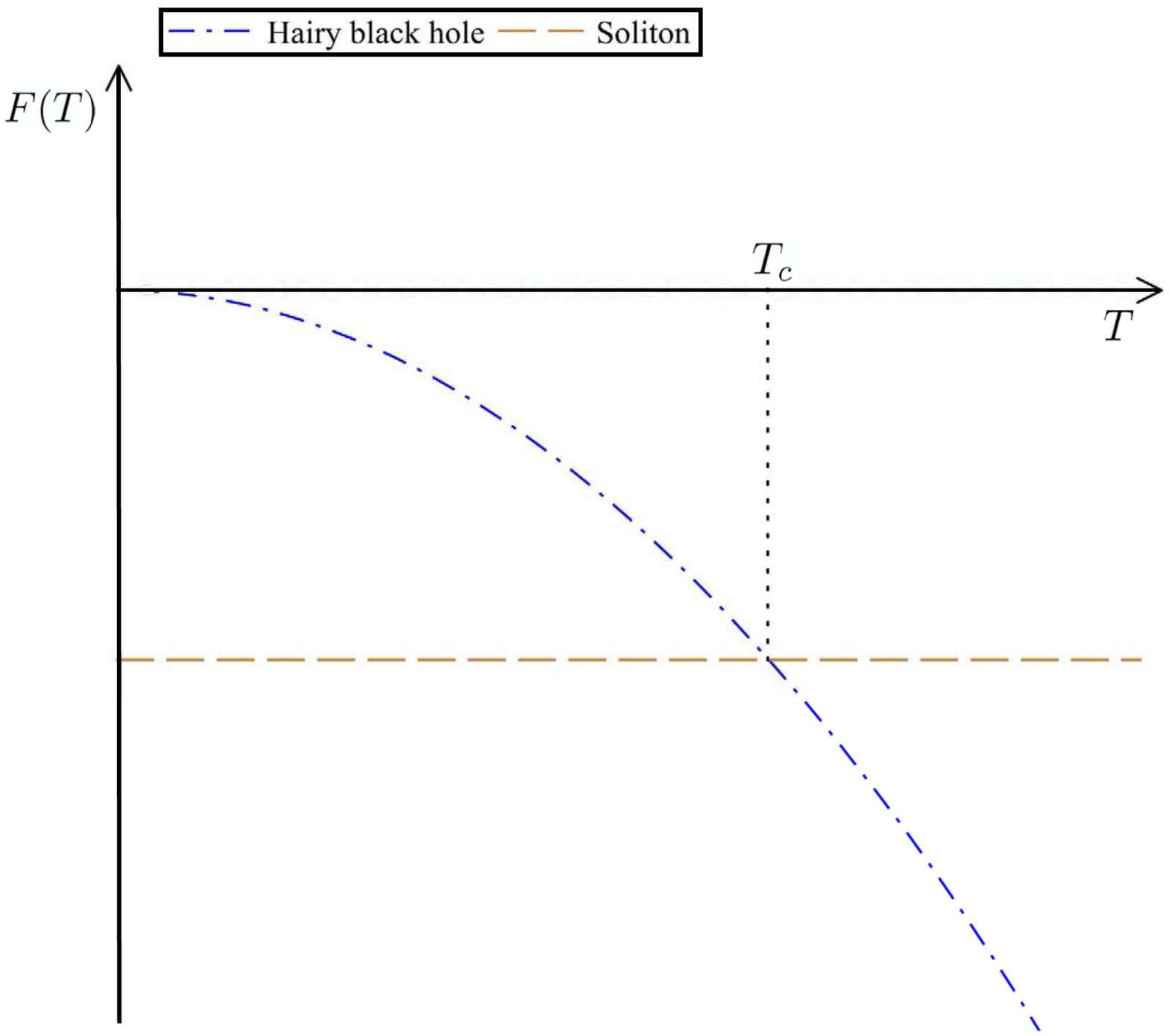}
\end{center}
\caption{Free energy as a function of the temperature for the soliton and the hairy black hole. Here $T_{c}=(2\pi l)^{-1}$.}
\label{fig:FreeEnergies}
\end{figure}
Note that there is another possible decay channel at fixed $\beta$,
corresponding to the extremal black hole. However, since its free energy
vanishes, the extremal hairy black hole then always becomes unstable against
thermal decay.

Since the first derivative of the free energy has a discontinuity at the
critical temperature (\ref{Tc}), given by%
\begin{equation}
\left.  \frac{\partial F(T)}{\partial T}\right\vert _{+}-\left.
\frac{\partial F(T)}{\partial T}\right\vert _{-}=-\frac{\pi l}{G}\ ,
\end{equation}
the phase transition between the hairy black hole and the soliton is of first
order. As shown below, the presence of gravitational hair induces an
additional effective length scale which might allow a suitable treatment of
this phase transition in the semiclassical regime.

\subsection{Gravitational hair, thermal fluctuations and phase transition in
the semiclassical regime}

\label{Thermal fluctuations}

In the static case the hairy black hole metric (\ref{Rot-BH}) acquires a very
simple form, given by\footnote{This metric was first found in the context of
conformal gravity in three dimensions \cite{Joao Pessoa}, and for BHT massive
gravity it was independently discussed in \cite{OTT} and \cite{BHT2}.}%
\begin{equation}
ds^{2}=-\frac{1}{l^{2}}\left(  r-r_{+}\right)  \left(  r-r_{-}\right)
dt^{2}+\frac{l^{2}}{\left(  r-r_{+}\right)  \left(  r-r_{-}\right)  }%
dr^{2}+r^{2}d\phi^{2}\ ,\label{BH metric}%
\end{equation}
where the integration constants, $r_{-}<r_{+}$, can be expressed in terms of
the mass and the hair parameter according to%
\begin{equation}
\mathcal{M}=\frac{1}{16Gl^{2}}\left(  r_{+}-r_{-}\right)  ^{2}\ ,\label{M}%
\end{equation}%
\begin{equation}
b=\frac{1}{2l}\left(  r_{-}+r_{+}\right)  \ .\label{bStatic}%
\end{equation}
The solutions splits in two branches according to the sign of $b$, such that
for $b<0$ there is a single event horizon located at $r=r_{+}$, provided the
mass parameter fulfills the bound (\ref{Mass bound}) with $a=0$. In the case
of $b>0$ the constants $r_{-}$ and $r_{+}$ correspond to the Cauchy and the
event horizons respectively. Hence, in this case the hair parameter introduces
an effective length scale that corresponds to the horizon radius of the
extremal black hole, given by $r_{+}=r_{-}=r_{e}$, with
\begin{equation}
r_{e}:=bl\ ,\label{Re}%
\end{equation}
that fixes the minimum size of the hairy black hole ($r_{+}\geq r_{e}$).

In order to explore the nature of the phase transition it is useful looking at
the behaviour of the specific heat $C=\frac{d\mathcal{M}}{dT}$, which for the
static hairy black hole (\ref{BH metric}) is given by%
\begin{equation}
C=\frac{2\pi^{2}l^{2}}{G}T=\frac{\pi}{G}\left(  r_{+}-r_{e}\right)  \ .
\label{C}%
\end{equation}
Since the specific heat (\ref{C}) is positive for a non extremal black hole,
it can reach local thermal equilibrium with a heat bath provided the thermal
fluctuations are small. Note that for the near extremal black hole,
$r_{+}\gtrsim r_{e}$, the heat capacity becomes arbitrarily small, and
therefore a fixed amount of energy that is either absorbed or radiated by the
black hole, necessarily implies a large fluctuation of the temperature. In
this sense, the near extremal black hole becomes \textquotedblleft
volatile\textquotedblright, which signals the existence of the phase
transition being triggered by thermal fluctuations. Indeed, the energy and
temperature fluctuations are given by\footnote{In the canonical ensemble the
fluctuations are related to the specific heat as $\left(  \Delta E\right)
^{2}=CT^{2}$, and $\left(  \Delta T\right)  ^{2}=C^{-1}T^{2}$.}%
\begin{equation}
\frac{\left(  \Delta E\right)  ^{2}}{E^{2}}=\frac{4}{\pi}\frac{\ell_{p}%
}{\left(  r_{+}-r_{e}\right)  }\ , \label{Efluct}%
\end{equation}%
\begin{equation}
\frac{\left(  \Delta T\right)  ^{2}}{T^{2}}=\frac{\ell_{p}}{\pi\left(
r_{+}-r_{e}\right)  }\ , \label{Tfluct}%
\end{equation}
respectively, where $\ell_{p}:=G$ is the Planck length. Since for a near
extremal black hole the difference $r_{+}-r_{e}$ is very small, according to
Eqs. (\ref{Efluct}) and (\ref{Tfluct}), in this case the energy and
temperature fluctuations become very large, and hence hairy black holes turn
out to be thermodynamically unstable at low temperatures.

Remarkably, due to the existence of gravitational hair, which fixes the size
of the extremal hairy black hole as in Eq. (\ref{Re}), the transition is able
to take place in the semiclassical regime of the theory which ensures the
reliability of the previous analysis. This is because in the semiclassical
approximation the event horizon, extremal, and AdS radii have to be much
larger that the Planck length, i.e., $l,r_{+},r_{e}\gg\ell_{p}$. This means
that the growth of the thermal fluctuations for a low temperature hairy black
hole can be seen occur in the semiclassical approximation, provided the scale
introduced by the gravitational hair, which fixes the radius of the extremal
black hole fulfills $r_{e}\gg\ell_{p}$. Note that this is not the case for the
BTZ black hole, for which $r_{e}=0$, because in the semiclassical regime
$r_{+}\gg\ell_{p}$ the fluctuations become very small, so that the possible
phase transition would occur in a regime where quantum gravity effects become
relevant (see e.g. \cite{CruzLepe,KuritaSakagami,Reznik,CaiLuZhang}).

As an ending remark, it is worth pointing out that a usual feature of first
order phase transitions, as it occurs for water, is the possibility of bubble
nucleation due to the existence of metastable states. Indeed, apart form the
critical temperature $T_{c}$ defined in Eq. (\ref{Tc}), it is useful to
introduce a temperature $T_{1}$, given by%
\begin{equation}
T_{1}=\frac{4}{\pi}\left(  \frac{\ell_{p}}{l}\right)  T_{c}\ ,\label{T1}%
\end{equation}
for which the energy fluctuations (\ref{Efluct}) are of order one. Thus, for
$T>T_{c}$ the black hole clearly dominates the partition function, while for
$T_{1}<T<T_{c}$ the black hole turns out to be metastable against vacuum
(soliton) nucleation, since it possesses less free energy than the soliton but
the thermal fluctuations are too small so as to trigger a sudden transition.
For $T<T_{1}$ the vacuum in a thermal bath of radiation is the preferred
configuration. Note that in the transition from the ground state to the black
hole, the size that characterizes an extremal black hole, determined by the
hair parameter, would be spontaneously chosen as the temperature increases.

\section{Summary and discussion}

\label{Discussion}

The purely gravitational soliton described by the metric
(\ref{Soliton nice coordinates}) was shown to possesses a fixed negative mass
$M_{\mathrm{sol}}$ given by (\ref{Msol}) which does not depend on the
integration constant $\alpha$. Thus, the soliton can be naturally regarded as
a degenerate ground state labeled by a single modulus parameter, whose mass
precisely coincides with the one of AdS spacetime. According to Eqs.
(\ref{Isol}) and (\ref{IBH1}), the Euclidean action (\ref{reg1-1}) was shown
to be finite and independent of modulus and hair parameters for the soliton as
well as for the hairy black hole (\ref{Rot-BH}). It is then amusing to verify
that the hair parameter $b$, an integration constant that cannot be gauged
away since it has an apparent effect in the causal structure structure of the
hairy black hole, has a missing role in the global charges. The corresponding
masses (\ref{Msol2}) and (\ref{Mass}), with the angular momentum
(\ref{Angular momentum}) and the entropy (\ref{BH-Entropy}) in the case of the
black hole were successfully recovered from a different method. It was also
shown that the Euclidean actions of the soliton and the rotating hairy black
hole can be written as in Eqs. (\ref{IsolTau}) and (\ref{IhbhTau}),
respectively, so that they are completely determined by the soliton mass
$M_{sol}$ and the modular parameter $\tau$ in Eq. (\ref{Modular parameter}),
as it occurs for a different class of black holes with scalar hair and scalar
solitons in General Relativity \cite{HMTZ-2+1, CMT-Soliton}. Both Euclidean
actions turned out to be related by a modular transformation given by
(\ref{Modular inv}). This is in full agreement with the case of Euclidean AdS
and the BTZ black hole in General Relativity \cite{Maldacena-Strominger},
which is a consequence of the fact that both Euclidean solutions are
diffeomorphic \cite{Carlip-Teitelboim}. In this sense, here the soliton plays
the role of AdS in General Relativity. Indeed, according to Eqs.
(\ref{ModularTransfMetrics1}) and (\ref{ModularTransfMetrics2}), the Euclidean
rotating black hairy hole becomes diffeomorphic to the Euclidean soliton,
provided the quotient of the modulus and hair parameter is fixed by the norm
of $\tau$ as in Eq. (\ref{Modulus-hair-map}), which remarkably maps the mass
bound required by cosmic censorship for the hairy black hole (\ref{Mass bound}%
) with the condition (\ref{Soliton existence condition}) that guarantees the
regularity of the soliton at the origin. The relationship
(\ref{Modulus-hair-map}) further suggests that the hairy black hole is in a
broken phase of the theory, in which the modulus parameter of the ground state
is spontaneously fixed. This is supported by the fact that Cardy formula
(\ref{Cardy super reloaded})\ agrees with the semiclassical entropy
(\ref{BH-Entropy}) when the degeneracy of the ground state is removed (see Eq.
(\ref{Number of states})). It is then reassuring to verify that there is a
critical temperature $T_{c}=(2\pi l)^{-1}$ characterizing a phase transition
between the static hairy black hole and the soliton. This phase transition is
of first order and it turns out to be qualitatively different than the one of
Hawking and Page \cite{HawkingAndPage} between the Schwarzschild-AdS black
hole and AdS spacetime (for a recent discussion see \cite{Hawking-Page-2nd
Order}). Moreover, the existence of gravitational hair parameter induces an
additional effective length scale that determines the minimum size of the
black hole (see Eq. (\ref{Re})) which allows a suitable treatment of this
phase transition in the semiclassical regime.

One may speculate that the spontaneous choice of the modulus parameter
$\alpha$, which determines the gravitational hair, could be related with some
sort of symmetry breaking mechanism. Indeed, at the special point
(\ref{special point}) the linearized graviton becomes partially massless,
possessing an additional gauge symmetry which makes it to possess only one
degree of freedom. This symmetry is certainly broken by the self interactions
around a generic configuration, but it may survive around certain classes of
solutions that includes the soliton. Preliminary results indicate that there
is an enhancement of gauge symmetries around certain configurations at the
full nonlinear level \cite{PTT2}, as it has been observed for different
classes of degenerate dynamical systems (see e.g. \cite{CS}). Simple examples
of this phenomenon exist in classical mechanics \cite{STZ}, for which the rank
of the symplectic form may decrease on certain regions within the space of
configurations, so that around certain special classes of solutions,
additional gauge symmetries arise and then the system losses some degrees of
freedom. Unusual mechanisms like this one are clearly out of the hypotheses of
the Coleman-Mermin-Wagner theorem \cite{Coleman-Mermin-Wagner} (see e.g.
\cite{Anninos-Harnoll-Iqbal}), which also appears to be circumvented in the
context of holographic superconductors in $1+1$ dimensions \cite{Ren,
Lashkari, Liu-Pan-Wang}.

\bigskip

\textit{Acknowledgments.} It is a pleasure to thank Pedro Alvarez, Fabrizio
Canfora, Francisco Correa and specially to Cristi\'{a}n Mart\'{\i}nez for many
useful and enlightening discussions. This work has been partially funded by
the Fondecyt grants N$%
{{}^\circ}%
$ 1085322, 1095098, 3110141, 3110122 and by the Conicyt grant ACT-91:
\textquotedblleft Southern Theoretical Physics Laboratory\textquotedblright%
\ (STPLab). The Centro de Estudios Cient\'{\i}ficos (CECs) is funded by the
Chilean Government through the Centers of Excellence Base Financing Program of Conicyt.

\end{document}